\documentclass{article}

\usepackage{PRIMEarxiv}

\usepackage[utf8]{inputenc} % allow utf-8 input
\usepackage[T1]{fontenc}    % use 8-bit T1 fonts
\usepackage{hyperref}       % hyperlinks
\usepackage{url}            % simple URL typesetting
\usepackage{booktabs}       % professional-quality tables
\usepackage{amsfonts}       % blackboard math symbols
\usepackage{nicefrac}       % compact symbols for 1/2, etc.
\usepackage{microtype}      % microtypography
\usepackage{lipsum}
\usepackage{fancyhdr}       % header
\usepackage{graphicx}       % graphics
\graphicspath{{media/}}     % organize your images and other figures under media/ folder

\usepackage{multirow}%
\usepackage{amsmath,amssymb,amsfonts}%
\usepackage{amsthm}%
\usepackage{mathrsfs}%
\usepackage[title]{appendix}%
\usepackage{xcolor}%
\usepackage{textcomp}%
\usepackage{manyfoot}%
\usepackage{algorithm}%
\usepackage{algorithmicx}%
\usepackage{algpseudocode}%
\usepackage{listings}%
\usepackage{longtable}%----->
\usepackage{rotating}
\usepackage{comment}
\usepackage{subcaption}
\usepackage{caption}
\usepackage{tikz}
\usepackage{setspace} % Add this line to use the setspace package

\title{Utilizing AI and Social Media Analytics to Discover Adverse Side Effects of GLP-1 Receptor Agonists
\thanks{\textit{\underline{Citation}}: 
\textbf{Bartal et al., Utilizing AI and Social Media Analytics to Discover Side Effects of GLP-1 Receptor Agonists. Pages.... DOI:000000/11111.}} 
}

\author{
    Alon Bartal, Kathleen M. Jagodnik \\
    The School of Business Administration \\
    Bar-Ilan University\\
    Ramat Gan, Israel\\
    \texttt{\{bartal.alon, kathleen.jagodnik\}@biu.ac.ill} \\
   \And
  Nava Pliskin \\
  Department of Industrial Engineering \& Management \\
  Ben-Gurion University of the Negev \\
  Beer-Sheva, Israel\\
  \texttt{pliskinn@bgu.ac.il} \\
  \And
  Abraham Seidmann\\
  Questrom School of Business \\
  Boston University\\
  Boston, MA, USA \\
  \texttt{avis@bu.edu} \\
}

\begin{document}
\maketitle

\begin{abstract}
Adverse side effects (ASEs) of drugs, revealed after FDA approval, pose a threat to patient safety. 
To promptly detect overlooked ASEs, we developed a digital health methodology capable of analyzing massive public data from social media, published clinical research, manufacturers’ reports, and ChatGPT. 
We uncovered ASEs associated with the glucagon-like peptide 1 receptor agonists (GLP-1 RA), a market expected to grow exponentially to \$133.5 billion by 2030. 
Using a Named Entity Recognition (NER) model, our method successfully detected 21 potential ASEs overlooked upon FDA approval, including irritability and numbness. 
Our data-analytic approach revolutionizes the detection of unreported ASEs associated with newly deployed drugs, leveraging cutting-edge AI-driven social media analytics.
It can increase the safety of new drugs in the marketplace by unlocking the power of social media to support regulators and manufacturers in the rapid discovery of hidden ASE risks.
\end{abstract}

% keywords can be removed

%------------------------------------------------------------------
\section{Introduction}
\label{sec:intro}
%------------------------------------------------------------------
%The Problems of Obesity \& Type 2 Diabetes
Over recent decades, obesity has surged as a global health crisis \cite{who-obesity}, 
with numbers nearly tripling since 1975, surpassing 650 million individuals globally, among 1.9 billion adults who are overweight \cite{who-obesity}. 
Obesity significantly elevates the risk of developing an array of health disorders, including type 2 diabetes (T2D), high blood pressure, heart disease, respiratory problems, joint problems, and gallbladder disease 
 \cite{cdc_diabetes, pearson2022variations, khand2020association}. 
The consequences of obesity are projected to consume more than 3\% of the gross domestic product in the United States (U.S.) in the coming decades \cite{okunogbe2022economic}.

Glucagon-like peptide 1 receptor agonists (GLP-1 RA), used to treat obesity and T2D, effectively reduce weight, improve comorbidities, and lower blood sugar levels \cite{moore2023glp}. 
Since the U.S. Food and Drug Administration (FDA) approved the first drug in this class, exenatide, in 2005, the use of GLP-1 RA such as dulaglutide (Trulicity), liraglutide (Victoza), and semaglutide (Ozempic, Rybelsus) has surged \cite{trujillo2021glp}.
In 2023, prescriptions spiked, driven by both T2D and obesity treatments \cite{pritchett-2022-trends}. 
Semaglutide's popularity, fueled by celebrity uses for weight loss, led to supply shortages \cite{rodriguez2023first, han2023public}.
However, GLP-1 RA also have documented Adverse Side Effects (ASEs), including gastrointestinal disorders, metabolic problems, and eye, renal, urinary, and nutritional disorders \cite{cabral2023gastrointestinal,zhang2023glp}. 
Because some drugs in this class were approved in recent years (e.g., Ozempic in 2017 \cite{novo-nordisk-ozempic-2017}), 
their complete ASE profiles remain under-characterized, since clinical trials, typically involving only a few thousand human subjects, often cannot detect rare ASEs or those with significant latent development \cite{berlin2008adverse}.
Moreover, clinical trials can overlook real-world scenarios, necessitating careful consideration of the relevance of their findings for larger and more diverse populations. 
For example, non-obese individuals who use GLP-1 RA for weight loss are not commonly represented in clinical trials. 
Limited knowledge about ASEs solely from clinical trials lead physicians to prescribe medications with incomplete information, risking patient safety.

Furthermore, some FDA-approved drugs have been withdrawn from the market when their ASEs became known only after widespread use \cite{ridings2013thalidomide,ponrartana2021safety,el2022selective}. 
Thalidomide, for example, was prescribed to women during pregnancy to treat nausea, but led to severe birth defects in infants whose mothers had taken the drug, ultimately resulting in its withdrawal from the market \cite{ridings2013thalidomide}.
Vioxx, which was approved by the FDA for the treatment of pain and inflammation associated with arthritis, was eventually withdrawn due to its being associated with increased risk of heart attack and stroke \cite{ortiz2004market}.

These historical pharmacological precedents motivate the need to add another layer of safety beyond the initial FDA approval. 
Our proposed digital health methodology can serve as an effective `early warning' sensor. 
The approach should continuously monitor users' communities following the introduction of any new drug to identify any previously unreported ASEs that may stem from patients' biology, unexpected drug interactions, or individual lifestyle choices.

Drug ASEs are already recorded in databases such as the FDA Adverse Event Reporting System (FAERS) \cite{fda-faers-2023}, 
the Side Effect Resource (SIDER) \cite{kuhn2016sider}, 
MEDLINE \cite{ding2001mining}, 
and Embase \cite{elsevier-embase-2023}. 
However, these databases suffer from improper indexing \cite{derry2001incomplete}
and incomplete reporting \cite{nugent2016computational}, 
with up to 86\% of ASEs unreported \cite{nugent2016computational}. 
Moreover, publishing bias favoring positive results, as well as practices that may conceal negative data in publications \cite{ekmekci2017increasing, mlinaric2017dealing, dickersin2011recognizing}, 
can lead to incomplete recording of ASEs. 
Additionally, the diversity of resources providing data about ASEs, the heterogeneous and incomplete nature of these data \cite{nugent2016computational}, 
and failure of patients to report side effects \cite{foster2008use} 
make it challenging to definitively characterize the ASEs of any drug \cite{nugent2016computational,zhang2013exploring}.
% social media

Social media can supplement these resources by allowing researchers to preemptively detect ASEs before regulatory intervention \cite{lee2021use}.
For example, social media analyses have been conducted to identify ASEs of medications for HIV \cite{godinez2023analysis,loosier2021reddit}, 
chemotherapy for cancer \cite{jenei2021experiences,zhang2018utilizing}, 
and buprenorphine-naloxone treatment for opioid use \cite{graves2022thematic}. 
Specifically, Reddit \cite{godinez2023analysis, graves2022thematic, jenei2021experiences, loosier2021reddit} 
and $\mathbb{X}$ (formerly Twitter) \cite{zhang2018utilizing, hsu2017mining, jiang2013mining} 
have been used to discover pharmacovigilance knowledge.
The Reddit social media platform enables anonymous accounts, encouraging users to share ASE experiences more honestly compared with other platforms \cite{anderson2015ask}, 
as anonymity often leads individuals to disclose information more freely \cite{luo2020self,schlosser2020self,suler2004online}.
Reddit also permits more substantial exchanges, compared with $\mathbb{X}$, which limits users to brief posts.
However, the briefer $\mathbb{X}$ posts can also provide significant information about ASEs reported by users \cite{back2023adverse,jiang2020mining, hsu2017mining, jiang2013mining, bian2012towards} 
and about other aspects of drug usage \cite{bremmer2023social,alvarez2021areas}.

Using social media analysis, Adjeroh et al. found that ASEs could be detected 3 to 35 months before an FDA alert was issued, depending on the drug \cite{adjeroh2014signal}.
A scoping review of social media use for pharmacovigilance \cite{lee2021use} 
reported that social media monitoring can facilitate the detection of labeling changes, black box warnings, or drug withdrawals 3 months to 9 years in advance.
Additionally, recent meta-analysis \cite{gao2022does} 
has shown that a higher prevalence of ASE discussions on social media is associated with a shorter time to drug recall \cite{gao2022does}.

The rising popularity of GLP-1 RA has led to a surge in online searching \cite{han2023public,tselebis2023further} 
and social media posting \cite{keating2023semaglutide} 
about them, offering valuable data to supplement existing pharmacovigilance sources \cite{back2023adverse}.
While social media posts do provide abundant information about GLP-1 RA, previous research has mainly focused on small-scale data \cite{basch2023descriptive,lennon2023can}. 
For instance, recent analysis of the top 100 results of TikTok videos tagged \#Ozempic revealed that most videos emphasized personal experiences and promoted the drug for weight loss, potentially contributing to increased demand \cite{basch2023descriptive}. 
That study stressed the importance of collaboration among healthcare professionals, regulatory bodies, and social media platforms to address challenges like drug shortages. 
Another small-scale study on TikTok compared 16 videos in total, highlighting how Ozempic users framed obesity as a disease, unlike discussants of traditional diet methods \cite{lennon2023can}.
Researchers have also started to use the abundant information that social media offers about GLP-1 RA for broader analysis.
For example, a recent large-scale study \cite{arillotta2023glp} 
analyzed tens of thousands of posts on Reddit (12,136), YouTube (14,515), and TikTok (17,059) and uncovered associations between GLP-1 RA and various mental health issues.
Our work responds to the authors' call for further study to capture subsequent comments after prolonged drugs use.

Training Natural Language Processing (NLP) models, including those utilizing Named Entity Recognition (NER), is integral to extracting knowledge from unstructured text sources such as social media posts, electronic medical records, and medical literature \cite{tarcar2019healthcare}.
NER plays a crucial role in identifying biomedical entities such as drugs, symptoms, and diseases mentioned in the text, with potential implications for linking them to ASEs \cite{tarcar2019healthcare}. 
Advanced NER models, leveraging deep-learning techniques \cite{li2020survey}, 
and utilizing contextual information surrounding entities within sentences, benefit from Large Language Models (LLMs) that are typically pre-trained on massive publicly and freely available text on the Internet, encompassing books, articles, and webpages \cite{sciencefocus-gpt3}.
LLMs possess the capability to receive and generate text queries while adapting to a variety of language-related tasks beyond their primary training objective \cite{omiye2024large}.

In the medical field, LLMs have been applied to tasks like responding to medical exams and addressing patient queries.
Despite the risk of generating inaccurate outputs, healthcare providers are increasingly integrating LLMs into clinical applications \cite{lee2023benefits}. 
Noteworthy examples include LLM use in medical resident training modules and in partnerships of healthcare providers that incorporate electronic health records \cite{statnews-chatgpt4-2023,healthcareitnews-epic-microsoft-2023}. 
Within the landscape of general LLMs, the Generative Pre-trained Transformers (GPT) lineage, particularly exemplified by OpenAI's ChatGPT, emerges as a frequently used choice, especially in chat applications \cite{omiye2024large}.

% ChatGPT
ChatGPT is an LLM known for generating informative text and mimicking human-like writing \cite{chen2023extensive, lancaster2023artificial}. 
It has been widely used in various scientific domains, including mental health research \cite{bartal2023identifying, lamichhane2023evaluation}, 
medical research \cite{lee2023ai, ruksakulpiwat2023using}, 
and detecting drug interactions \cite{juhi2023capability}. 
However, ChatGPT's capabilities have not yet been used to study GLP-1 RA, presenting a valuable opportunity for knowledge discovery through the methods harnessed in the present study.

Computational-based algorithms have already facilitated understanding of drug ASEs, including identifying when they occur \cite{mcmaster2023developing, trajanov2023review}.
Accurate assessment of ASE frequencies is crucial for patient care in clinical practice.
Drug-related ASEs persist as a significant cause of morbidity and mortality in healthcare, resulting in substantial economic costs, amounting to an estimated annual cost of \$495-672 billion in the U.S. alone \cite{watanabe2018cost}.
Moreover, accurate estimation can help pharmaceutical companies to mitigate the risk of a drug's withdrawal from the market \cite{onakpoya2016worldwide} 
and avoid the need for costly reassessment of ASEs through new clinical trials \cite{martin2017much}.

This study contributes to the understanding of ASEs related to GLP-1 RA through a multifaceted knowledge-discovery methodology.
Our approach mines data from social media, extending beyond existing pharmacovigilance insights provided by drug manufacturers and clinical trials, to comprehensively identify and analyze the ASEs of GLP-1 RA. 
Our construction of an ASE-ASE network, based on co-mentioned ASEs in social media posts, represents a novel approach to revealing groups of ASEs with similar network patterns.
We further contribute by applying network community detection and classification based on mention frequency, offering a nuanced perspective on ASE prevalence and interconnections. 
Overall, our study contributes methodologically and conceptually by offering valuable insights for drug safety assessment, public health management, and the broader understanding of GLP-1 RA related ASEs.

The rest of this paper is organized as follows:
Section \ref{sect:RQ} elaborates on the research question and hypotheses. 
Section \ref{sec: material} provides a detailed description of the methodology employed.
It also presents the characteristics of the data sources we used to produce Section \ref{sec:res}'s results.
Finally, the discussion and conclusions in Section \ref{sec:discussion} interpret the findings in the context of the existing literature, explore their implications, and summarize key insights that suggest future research.

%------------------------------------------------------------------
\section{Research Questions and Hypotheses}
\label{sect:RQ}
%------------------------------------------------------------------

\textbf{RQ1.} How can we identify unreported ASEs of GLP-1 RA based on social media discussions?

Health-focused individuals frequently use social media platforms to seek advice and exchange experiences with others, especially concerning ASEs \cite{ozurumba2022multi,chen2021social}.
These textual posts can be mined using NLP models.
Thus, we hypothesize:

\textit{H1.} A Named Entity Recognition (NER) model applied to social media posts can uncover ASEs of GLP-1 RA that are not observed by traditional pharmacovigilance methods.

%------------------------------------------------------------------
\section{Materials and Methods}
\label{sec: material}
%------------------------------------------------------------------
\subsection{Data}
\label{sect:data}
%------------------------------------------------------------------
We collected five datasets covering biomedical knowledge and social media posts involving the GLP-1 RA that are included on our 12-item \textit{Medication List}, including both brand names and generic drug names:
    dulaglutide (brand name: Trulicity),
    exenatide (brand names: Byetta; Bydureon - extended-release),
    liraglutide (brand name: Victoza), 
    lixisenatide (brand name: Adlyxin), and
    semaglutide (brand names: Ozempic - injections; Rybelsus - tablets).

\textbf{Dataset 1: $\mathbb{X}$ --}
Using the APIfy\footnote{https://console.apify.com} website, we collected the 1,000 most recent posts from 2017 to 2023 that mentioned each of the 12 items on our \textit{Medication List}.
This produced a dataset that includes 11,185 posted texts, 
each with associated data on the posting user, posting time, and GLP-1 receptor agonist mentioned.

\textbf{Dataset 2: Reddit --}
Using the Python Reddit API Wrapper (PRAW) package \cite{boe2023praw}, 
we collected 489,529 posts and comments from 14 subreddits (Table \ref{tbl:reddit}) related to the 12 items on our \textit{Medication List}.
The posts originated in 2022 and 2023.
Each collected Reddit post includes the following metadata:
    `Post ID',
    `Post Author',
    `Post Content',
    `Post Date',
    `Comment ID',
    `Comment Author',
    `Comment Content',
    `Parent ID', and
    `Parent Author'.

\begin{table}[h]
\caption{Subreddits that were used to collect social media posts involving GLP-1 RA.}
\label{tbl:reddit}
\centering
\begin{tabular}{ll}
\toprule
Subreddit               & \# Members$^*$ \\
\midrule
r/diabetes            & 109K \\
r/diabetes\_t2            & 29.4K \\
r/GLP1       & 1.3K \\
r/liraglutide            & 11.7K \\
r/loseit         & 3900K \\
r/MaintenancePhase             & 25.6K \\
r/medicine           & 453K \\
r/Ozempic                  & 50.1K \\
r/OzempicForWeightLoss     & 13.7K \\
r/semaglutidecompounds           & 7.2K \\
r/Semaglutide              & 45K \\
r/TheMorningToastSnark              & 11.6K \\
r/trulicity              & 1.1K \\
r/type2diabetes           & 8.4K \\
\bottomrule
\end{tabular}
\\
$^*$Number of members (in thousands) of each subreddit as of October 2, 2023.
\end{table}

\textbf{Dataset 3: PubMed.}
Using the National Center for Biotechnology Information (NCBI) PubMed E-utilities API (PubMed API) \cite{sayers2023eutilities} 
and the Biopython Python library \cite{chapman2000biopython}, 
we collected 13,491 articles indexed from 2017 to 2024 in PubMed that contained at least one item on our \textit{Medication List} in the title or in the abstract of the article.

\textbf{Dataset 4: SIDER (Side Effect Resource).}
We utilized the \texttt{meddra\_all\_se.tsv.gz} dataset from the SIDER database version 4.1, which catalogs 5,868 documented ASEs for 1,430 commercially available drugs. 
SIDER compiles this information from publicly accessible drug documents and package inserts, providing details on drug names, ASE frequencies, and classifications.

\textbf{Dataset 5: Side Effects Reported by Manufacturers and by ChatGPT.}
We manually searched on the web to collect ASEs reported by drug manufacturers, whose reports primarily include structured data from clinical trials and post-marketing surveillance.
Table \ref{tbl:manu} shows the latest update dates for GLP-1 RAs' ASEs by manufacturers. 
We recorded all prominent ASEs reported in the manufacturers' documents, focusing on data recorded from clinical trials with human subjects, and omitting side effects reported only in animal models.
To address potential gaps in the manually searched ASE data reported by manufacturers, we integrated them into Dataset 5 with commonly reported ASEs that we collected from ChatGPT (GPT-3.5) by using the interactive web interface. 
For this collection, we utilized the prompt: `Please provide a list of adverse side effects for the GLP-1 receptor agonist named $S$,'  where $S$ corresponds to an item on our GLP-1 RA \textit{Medication List}. 
ChatGPT's responses may incorporate a broader range of information, potentially including insights from social media or medical literature, that is not explicitly included in our manual search of manufacturers' reports. 
This approach ensures that Dataset 5 aligns more closely with the timeframes of Datasets 1 and 2, covering information up to and including 2023, and Dataset 3, which covers information up to and including January 2024.
  
\begin{table}
\caption{Last update dates and side effect URLs for GLP-1 RA}
\label{tbl:manu}
\centering
\begin{tabular}{|l|l|l|p{6cm}|}
\hline
\textbf{Drug Name} & \textbf{Brand Name} & \textbf{Last Update} & \textbf{Side Effects URL} \\
\hline
Exenatide & Byetta & 2009 & \url{https://www.accessdata.fda.gov/drugsatfda_docs/label/2009/021773s9s11s18s22s25lbl.pdf} \\
Lixisenatide & Adlyxin & 2016 & \url{https://www.accessdata.fda.gov/drugsatfda_docs/label/2016/208471orig1s000lbl.pdf} \\
Dulaglutide & Trulicity & 2017 & \url{https://www.accessdata.fda.gov/drugsatfda_docs/label/2017/125469s007s008lbl.pdf} \\
Exenatide & Bydureon & 2017 & \url{https://www.accessdata.fda.gov/drugsatfda_docs/label/2017/209210s000lbl.pdf} \\
Semaglutide & Ozempic & 2017 & \url{https://www.accessdata.fda.gov/drugsatfda_docs/label/2017/209637lbl.pdf} \\
Liraglutide & Victoza & 2019 & \url{https://www.accessdata.fda.gov/drugsatfda_docs/label/2019/022341s031lbl.pdf} \\
Semaglutide & Rybelsus & 2019 & \url{https://www.accessdata.fda.gov/drugsatfda_docs/label/2019/213051s000lbl.pdf} \\
\hline
\end{tabular}
\end{table}

%------------------------------------------------------------------
\subsection{Methods}
\label{sect:methods}
%------------------------------------------------------------------
In Sections \ref{sec:ASE} and \ref{sec:modelval}, we describe the method used to identify ASEs of GLP-1 RA in Datasets 1 to 3, and to validate the identified ASEs against Dataset 4 in comparison with Dataset 5.
In Section \ref{sec:mt}, we study the frequency of ASE mentions on social media.
Finally, in Section \ref{sec:cluster_side_effect}, we describe the method employed to construct an ASE-ASE network, which is based on co-mentions of ASEs within the same social media post (Datasets 1 and 2), and detail our approach to detecting ASEs that tend to co-occur by clustering ASE nodes in an ASE-ASE network based on the frequency of their mentions in the posts.

\begin{figure}
    \centering
    \includegraphics[scale=0.56, trim={0cm 0cm 0cm 0cm}, clip]{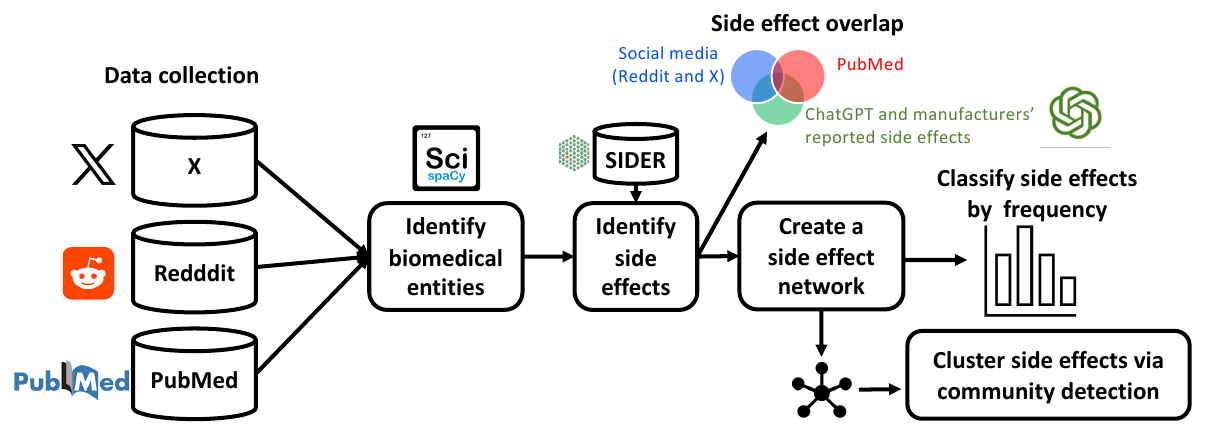}
    \caption{Overview of the methodology of this research.}
    \label{fig:methods}
\end{figure}

Fig. \ref{fig:methods} summarizes the steps of our approach to address the two research questions.

%------------------------------------------------------------------
\subsubsection{Identification of Adverse Side Effects} 
\label{sec:ASE}
%------------------------------------------------------------------
To identify ASEs of FDA-approved GLP-1 RA, we analyzed the texts in Datasets 1 to 3 and extracted biomedical entities, which included ASEs.
This process used the ScispaCy pre-trained NER model \cite{neumann2019scispacy}, 
specifically employing the \texttt{en\_ner\_bc5cdr\_md} NER model to identify biological entities.
This approach follows the strategy of utilizing ScispaCy for extracting pharmaceutical-related phrases, including dosage, diseases, and symptoms from electronic medical records \cite{tarcar2019healthcare}.
To ascertain whether these identified biological entities are potential ASEs of GLP-1 RA, we matched them against the ASEs catalogued in SIDER \cite{kuhn2016sider}, 
which were collected in Dataset 4.
This involved normalizing the potential ASEs and those in SIDER to lowercase and employing NLP stemming to reduce words to their root forms \cite{jivani2011comparative}.
Each entity was considered a potential ASE if it existed in SIDER \cite{kuhn2016sider}.
To the resulting compilation of ASEs, we added ASEs collected in Dataset 5 from manufacturers’ reports and ChatGPT. 
Then, we grouped similar ASEs, for example, combining `headache' and `migraines' and standardized ASEs into their noun forms.
Consolidating these diverse sources facilitated comprehensively identifying ASEs.

%------------------------------------------------------------------
\subsubsection{Method Evaluation: Identify Unreported ASEs}
\label{sec:modelval}
%------------------------------------------------------------------
We examined all ASEs identified in the social media data and classified them into one of two group:
\begin{itemize}
    \item 
    Established ASEs: ASEs reported by manufacturers and ChatGPT.

    \item 
    Novel ASEs: ASEs identified by our modeling approach but not reported by manufacturers and ChatGPT. 
\end{itemize}

If our method is able to identify established ASEs, we can confidently explore novel ASEs that we discovered and hypothesize their novelty.
Therefore, we evaluated our method's performance in identifying established ASEs by defining the Overlap score (Equation 1):

\begin{equation}
Overlap(f_\%)=|P_{f_\%} \cap I_d |/|I_d|
\end{equation}

$P_{f\%}$ - The set of established and novel ASEs identified in social media data located in the top $f_\%$ of a ranking, based on the mentioned frequency of ASEs.

$I_d$ - The set of established ASEs.

A higher Overlap score indicates greater success identifying established ASEs. 
A maximum Overlap of 1 indicates that all established ASEs were captured on social media, whereas a minimum Overlap of 0 indicates that no established ASEs were captured.

%------------------------------------------------------------------
\subsubsection{Analysis of Adverse Side Effects Mention Frequency}
\label{sec:mt}
%------------------------------------------------------------------
Addressing RQ1, which focuses on identifying novel ASEs based on social media data, we effectively separate and detail the frequency of ASE mentions on social media for each specific GLP-1 RA.
Additionally, we tracked mentions of ASEs on social media along intervals of 14 days.
To measure ASE frequency over time, we defined two frequency measures, considering only ASEs with more than ten mentions per interval:

\begin{itemize}
\item
    Pre-mention average frequency (Pre-MAF): average mentions before a selected point in time. 
\item 
    Post-mention average frequency (Post-MAF): average mentions after a selected point in time.
\end{itemize}

We define a positive slope in mention frequency as Pre-MAF $<$ Post-MAF, a negative slope as Pre-MAF $>$ Post-MAF, and no slope otherwise.

Furthermore, we queried Google Trends \cite{woloszko2020tracking} 
for each specific GLP-1 RA on our \textit{Medication List}.
This web service offers insights into how frequently a particular term is searched relative to the total search volume on Google.
In Google Trends the highest score, 100, represents the peak popularity of a particular term during the specified period, and the scores for other items are relative to this peak.

%------------------------------------------------------------------
\subsubsection{Analysis of Adverse Side Effects}
\label{sec:cluster_side_effect}
%------------------------------------------------------------------
Using the ASEs identified in the social media posts of Datasets 1 and 2 (Section \ref{sec:ASE}), we built an ASE-ASE network denoted by $G=(V,E,W)$. 
In $G$, nodes ($V$) are identified ASEs. 
Two nodes mentioned in the same post are connected by an edge ($E$). 
The weights of the edges ($W$) represent the frequency of two ASEs being mentioned together in posts.

Identifying the co-occurrence patterns of ASEs, which is crucial in pharmacovigilance drug safety analysis \cite{galeano2020predicting}, 
was accomplished in two steps.
We first investigated ASE-ASE clustering in $G$ using a community detection algorithm, as detailed below. 
Next, to improve risk assessment, inform treatment decisions, and contribute to overall pharmacovigilance and public health management, we classified ASEs by the frequency with which each was mentioned.

%------------------------------------------------------------------
\textbf{Clustering of ASEs by network community detection.}
%------------------------------------------------------------------
To identify sets of interconnected ASEs, we applied the \texttt{cluster\_louvain} node-clustering algorithm from the \texttt{igraph} R library to ASE-ASE network $G$. 
This algorithm employs a multi-level modularity optimization function, leveraging the modularity measure and adopting a hierarchical methodology.
Clustering nodes into community groups enabled the identification of ASEs that frequently co-occur beyond a direct 1-hop edge in $G$.

%------------------------------------------------------------------
\textbf{Classification of ASEs by mention frequency.}
%------------------------------------------------------------------
We also classified ASE nodes by the frequency with which each was mentioned in the social media posts of Datasets 1 and 2. 
Specifically, each ASE node was classified into one of the following groups (as defined in \cite{galeano2020predicting}): 
Very Rare, 
Rare, 
Infrequent, 
Frequent, and 
Very Frequent. 
This labeling approach was used to estimate the ASE frequency.

%------------------------------------------------------------------
\section{Results}
\label{sec:res}
%------------------------------------------------------------------
In Section \ref{subsec:res_ASE}, we identify ASEs and evaluate our method's performance in identifying established and novel adverse side effects (ASEs).
Then, In Section \ref{sec:freq}, we analyze the frequency of ASE mentions across social media (Datasets 1 and 2), academic literature (Dataset 3), and manufacturer and ChatGPT reports (Dataset 5), exploring temporal trends and correlations with external events.
Finally, in Section \ref{subsec:ASE_net}, we uncover clusters of interconnected ASEs and their mention frequencies.

%------------------------------------------------------------------
\subsection{Identify Adverse Side Effects of GLP-1 RA} 
\label{subsec:res_ASE}
\begin{figure}
    \centering
    \includegraphics[scale=0.7, trim={0cm 0cm 0.4cm 0cm}, clip]{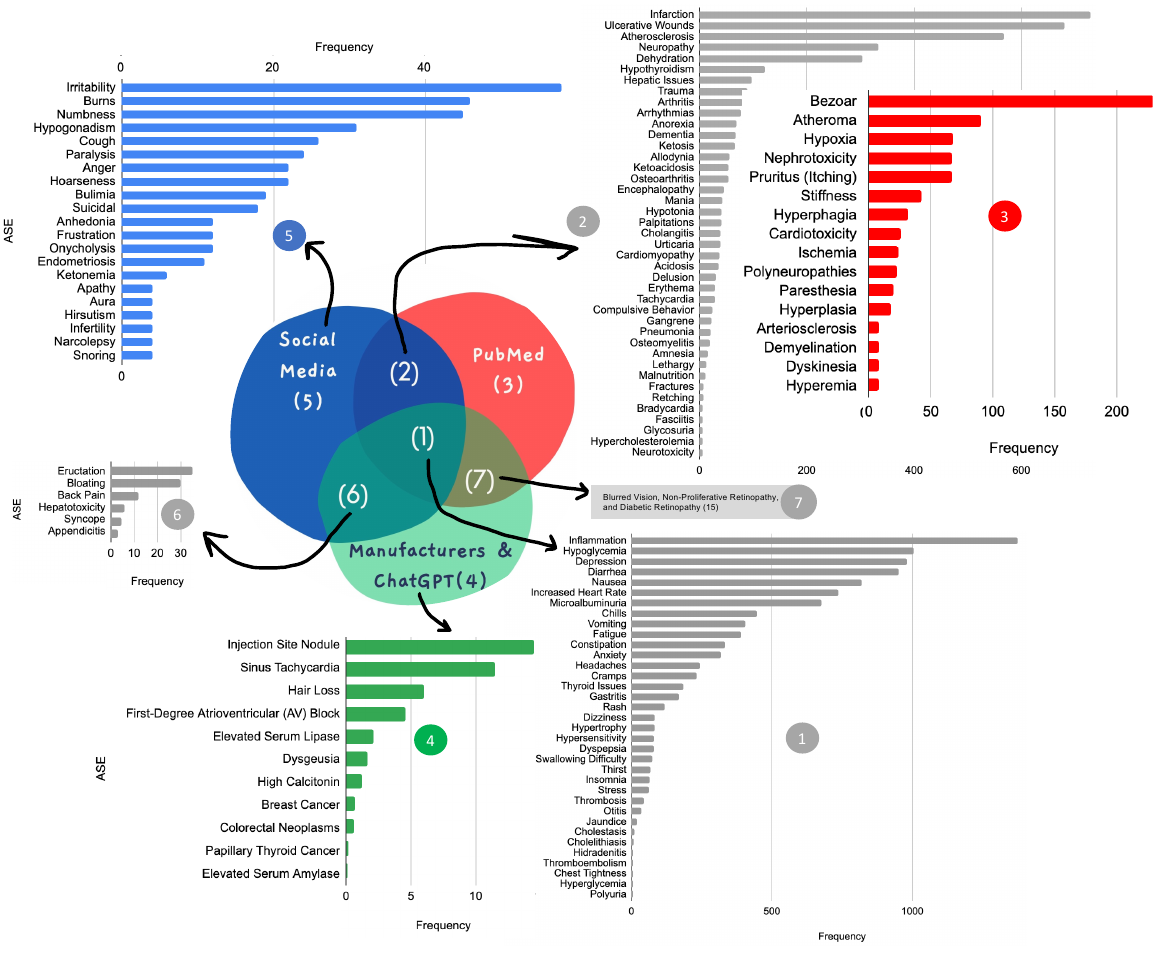}
    \caption{Venn diagram offering a visual representation of and distinction among ASEs.
    }
    \label{fig:venn}
\end{figure}

%------------------------------------------------------------------
Using the methods described in Section \ref{sect:methods}, we identified 134 ASEs of GLP-1 RA (Appendix Table \ref{app1:All_ASEs}).
Fig. \ref{fig:venn} presents a Venn diagram showing the overlaps and distinctions among these ASEs.
The bar charts in Fig. \ref{fig:venn} indicate ASE mention frequency in social media (Datasets 1, 2) and academic papers (Dataset 3), as well as ASE percentage frequency as reported by drug manufacturers (Dataset 5).
There are seven subgroups in the Venn diagram (intersection, union, and disjoint):
Group 1 lists ASEs found in academic papers, social media, and manufacturers' reports or reported by ChatGPT.
Group 2 lists ASEs reported in academic papers and mentioned on social media.
Group 3 lists ASEs reported exclusively in academic papers.
Group 4 lists ASEs reported by ChatGPT and in manufacturers’ reports. 
Group 5 lists ASEs mentioned exclusively on social media.
Group 6 lists ASEs mentioned on social media, by ChatGPT, and in manufacturers’ reports. 
Group 7 lists ASEs mentioned in academic papers, by ChatGPT, and in manufacturers’ reports. 
The frequencies of ASEs in Group 7 are summed to 15, as they describe similar ASEs: Blurred Vision, Non-Proliferative Retinopathy, and Diabetic Retinopathy.

%-----------------------------------------------------------------------
\subsubsection{Model Evaluation: Identify Novel ASEs}
\label{sec:model_eval}
%-----------------------------------------------------------------------
Following Section \ref{sec:modelval}, established ASEs are listed in Groups 1, 4, 6, and 7 in Fig. \ref{fig:venn}, and novel ASEs are listed in Groups 2 and 5.

Fig. \ref{fig:overlap} indicates the success of our model's identification of established ASEs as a function of $P_{f\%}$ ranging between 10\% to 100\% with increments of 10\%.
More specifically, it shows that the Overlap score increases with $P_{f\%}$.
For example, when $P_{f\%}=100\%$, indicating inclusion of all ASEs identified on social media, Overlap $= 53\%$ shows that our model identified $53\%$ of established ASEs. 
Similarly, in a recent study \cite{zitnik2018modeling}, a model was developed to predict drug-drug ASEs, with evidence found for 50\% of the identified ASEs in the literature.

\begin{figure}
    \centering
    \includegraphics[scale=0.4, trim={2.5cm 2.5cm 2.3cm 2.3cm}, clip]{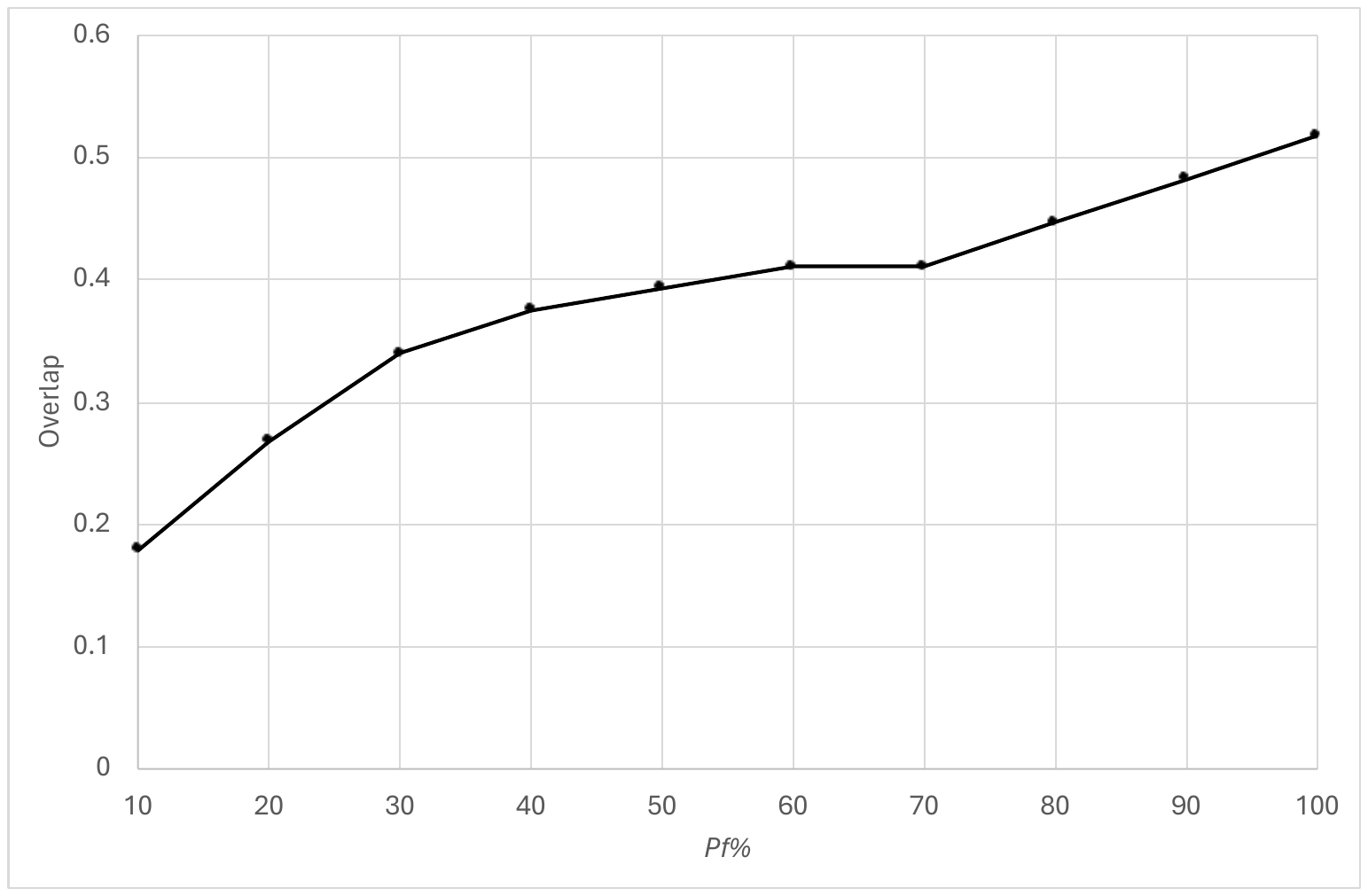}
    \caption{Overlap score as a function of $P_{f\%}$.
    }
    \label{fig:overlap}
\end{figure}

%-------
Next, following the identification of novel ASEs, we analyzed their frequency.

%------------------------------------------------------------------------------
\subsection{Analysis of ASE Mention Frequency}
\label{sec:freq}
%------------------------------------------------------------------------------
Following Section \ref{sec:mt}, Fig. \ref{fig:drg_ase_35} shows that Ozempic, which is the most frequently prescribed GLP-1 RA, has the largest group of associated ASEs.

\begin{figure}
    \ContinuedFloat
    \centering
    \begin{subfigure}[b]{\textwidth}
        \centering
        \includegraphics[scale=0.5, trim={0cm 0cm 0cm 0cm}, clip]{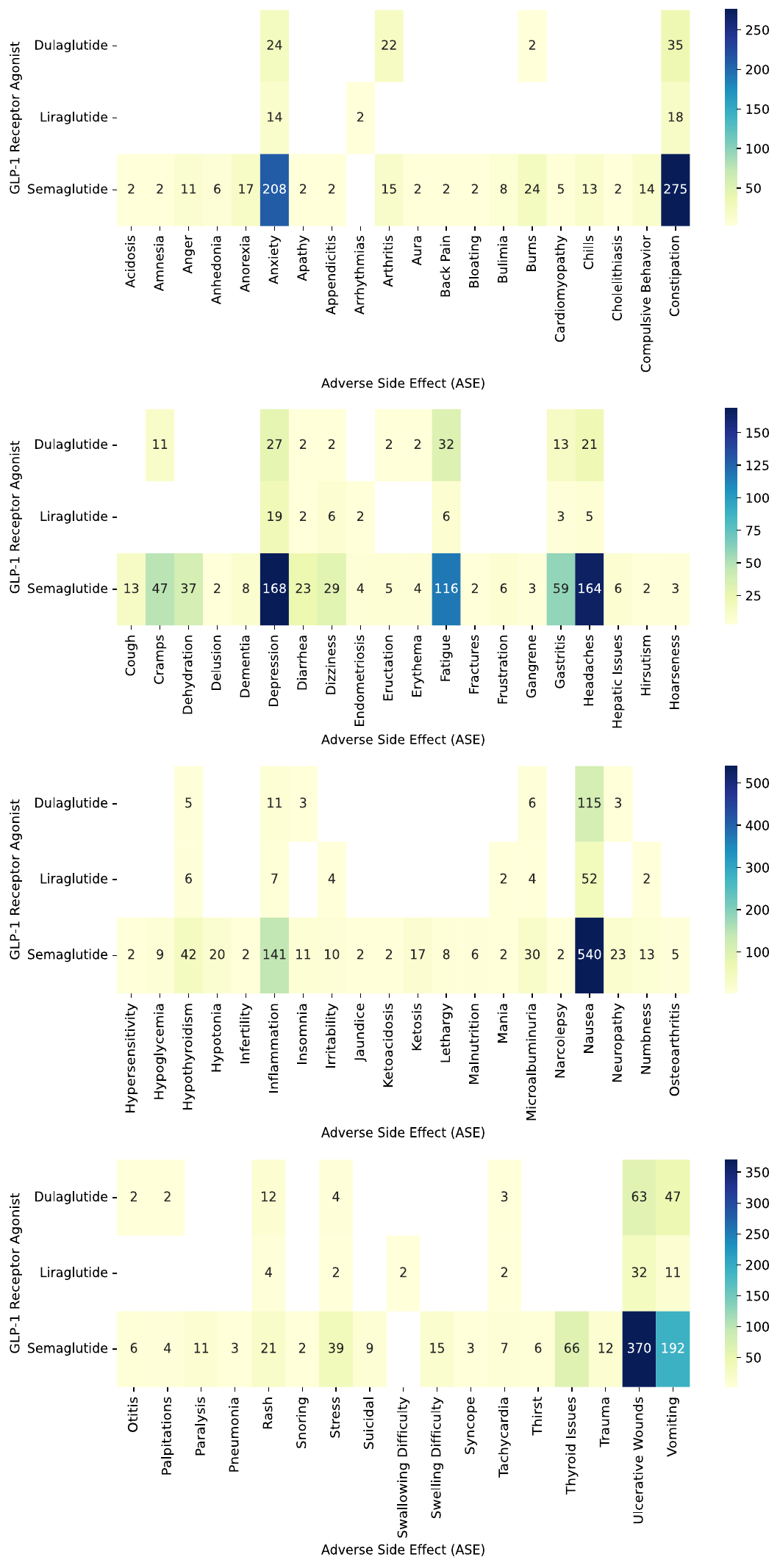}
        %\caption{Part B}
        \label{fig:drg_ase_35a}
    \end{subfigure}
    \stepcounter{figure} % Manually increment the figure counter
    \caption{ASE mention frequency ($>1$) on $\mathbb{X}$ and Reddit for each GLP-1 receptor agonist.}
    \label{fig:drg_ase_35}
\end{figure}

\begin{figure}
    \centering
    \includegraphics[scale=0.9, trim={7cm 3cm 10cm 7cm}, clip]{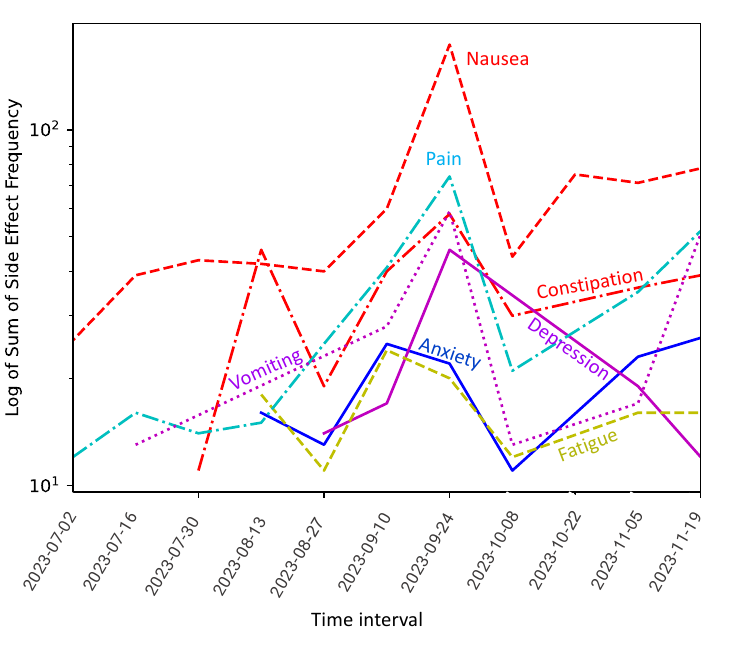}
    \caption{ASEs mentions $> 10$ in each of the 14-day intervals on $\mathbb{X}$ and Reddit. 
       }
    \label{fig:ASE_over_time}
\end{figure}

Tracking mentions of ASEs on social media along intervals of 14 days (Fig. \ref{fig:ASE_over_time}), we found that the ASEs mentioned most frequently over time are constipation, nausea, pain, and vomiting. 
We observed a sharp spike in these ASEs starting on September 24, 2023.
This date was selected to measure ASE frequency over time.
We observe an overall positive slope (Pre-MAF $<$ Post-MAF) for anxiety, constipation, nausea, pain, and vomiting, an overall negative slope for fatigue, and no slope for depression (Table \ref{tbl:freq_before_after}).

\begin{table}[h]
    \centering
    \caption{Frequency of adverse side effect (ASE) mentions before and after the observed spike in mentions.}
    \begin{tabular}{llll}
    \toprule
                Adverse Side Effect   & Pre-MAF & Post-MAF & Slope\\
                \midrule
                Anxiety & 18.0& 20.0 & +\\
                Constipation& 29.0& 35.0 & +\\
                Depression& 15.5& 15.5 & 0 \\
                Fatigue& 17.6& 14.6 & -\\
                %Inflammation& 12.0& NA\\
                Nausea& 39.1& 57.8 & +\\
                Pain (abdominal pain, back pain) & 20.5& 36.0 & +\\
                Vomiting& 20.5& 27.0 & +\\
                %Migraine& NA& 12.0\\
    \bottomrule
    \end{tabular}\\
    \label{tbl:freq_before_after}
\end{table}

\begin{figure}
    \centering
    \includegraphics[scale=0.5, trim={2cm 2.1cm 2cm 4cm}, clip]{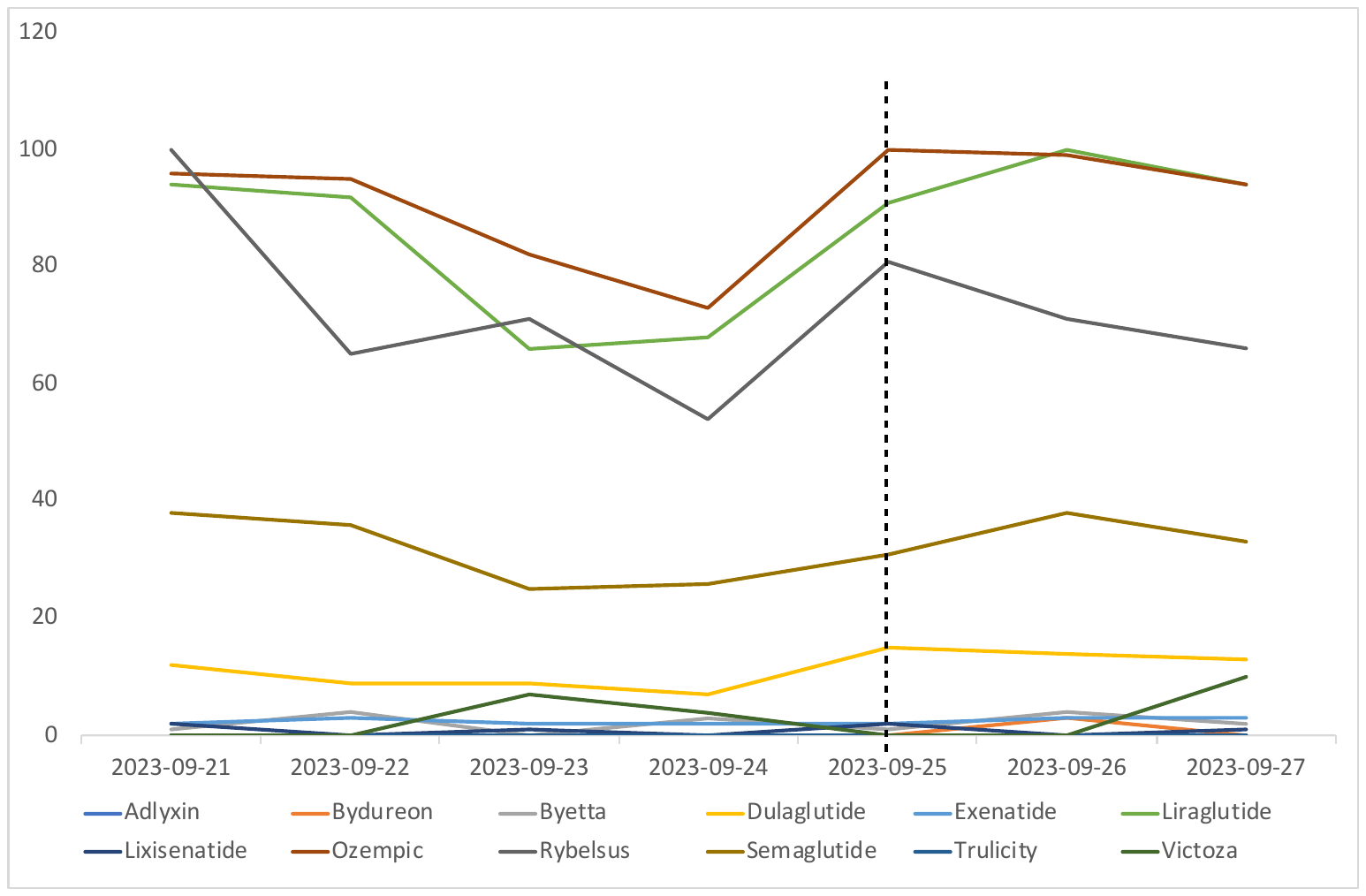}
    \caption{Google Trends queries for GLP-1 RA on our \textit{Medication List}.
        } 
    \label{fig:Gtrends}
    \begin{tikzpicture}[remember picture,overlay]
        \draw[black, thick, dashed] (1.95, 3.0) -- (1.95, 8.6); % Adjust the coordinates as needed
    \end{tikzpicture}
\end{figure}

To better understand the spike, we queried Google Trends \cite{woloszko2020tracking} 
for the entries on our \textit{Medication List} (Fig. \ref{fig:Gtrends}) with geographic location, category, and time range specified as `Worldwide', `Web Searches', and `Between 9/21/23 and 9/27/23' (i.e., 3 days before and after the 9/24/23 spike in Fig. \ref{fig:ASE_over_time}); we also filtered for `health'-related searches. 
Similar to Fig. \ref{fig:ASE_over_time}, Fig. \ref{fig:Gtrends} shows a spike on September 25, 2023, for the search terms `Dulaglutide', `Ozempic', `Liraglutide', `Trulicity', and `Rybelsus'.
The black dashed line in Fig. \ref{fig:Gtrends} represents a spike (Table \ref{tbl:freq_before_after}; Pre-MAF $<$ Post-MAF) in searches for `Dulaglutide', `Ozempic', `Liraglutide', `Trulicity', and `Rybelsus' on September 25, 2023.
The Google Trends delay of one day in the spike observed on September 24 may be attributed to the tendency of social media discussions to peak before gaining widespread attention through Google searches.
This temporal offset could arise from social media users acting as early adopters of a particular GLP-1 RA or possessing particular interest in a specific GLP-1 RA, thereby influencing the timing difference between the two spikes.

A further Google search for the analyzed time range helped to explain the spike by revealing that on her September 22, 2023, show, Oprah Winfrey, together with a panel of experts, addressed the growing trend of using weight-loss drugs.
She shared her initial reluctance to use such drugs, emphasizing the societal pressure to rely on personal effort rather than medication. 
The panel challenged the notion of weight loss solely relying on willpower and highlighted the roles of genetics and individual differences.

%---------------------------------------------------
\subsection{Analysis of Adverse Side Effects}
\label{subsec:ASE_net}
%------------------------------------------------------------------
Following Section \ref{sec:cluster_side_effect}, based on the unprocessed social media data (Datasets 1 and 2), we constructed an ASE-ASE weighted and non-directed network $G=(V,E,W)$.
To emphasize that our method can identify ASEs in raw social media discussions (RQ1), unlike in Section \ref{subsec:res_ASE}, we neither combined similar ASEs nor standardized ASEs into noun forms here.

\begin{figure}
    \centering
    \includegraphics[scale=0.8, trim={8cm 3cm 7cm 3.5cm}, clip]{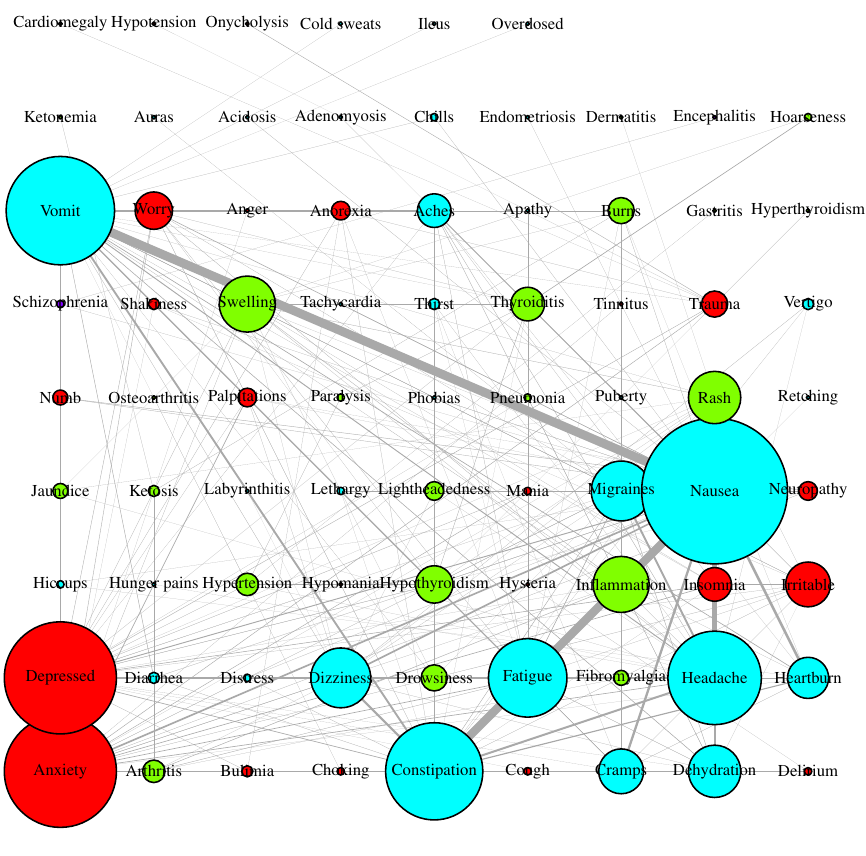}
    \caption{The ASE-ASE network $G$ based on social media posts.
    }
    \label{fig:graph}
\end{figure}

We identified 381 potential ASEs with 1,440 ASE-ASE interactions co-mentioned in the same post. 
To reduce noise and focus on ASEs only, we manually removed 52 falsely identified ASEs, ASE-ASE interactions with fewer than 3 co-mentions, and ASE nodes with no co-mentions in the data (having degree centrality 0). 
We also used the \texttt{components} function within the \texttt{igraph} R library \cite{csardi2013package} 
to detect graph components.
The algorithm detected three components of sizes 2, 2, and the main component of 78 nodes.
We set $G$ as the main component graph with 78 ASE nodes and 253 weighted edges.
In Fig. \ref{fig:graph}, nodes are ASEs and edges represent ASE-ASE co-mentioned relationships identified in $\mathbb{X}$ and Reddit posts.  

\textbf{Clustering of ASE nodes by network community.}
To identify sets of interconnected ASEs, we applied clustering analysis to $G$ using the \texttt{cluster\_louvain} node-clustering algorithm \cite{held2016dynamic}.
We discovered four clusters (Fig. \ref{fig:graph}) with 34, 23, 19, and 2 ASE nodes, respectively.
The node colors in Fig. \ref{fig:graph} correspond to membership in each of the four clusters; node size is proportional to its degree, and edge width is proportional to its weight, representing the co-mention frequency.
In Fig. \ref{fig:graph},
Cluster 1 (light blue) contains aches, acidosis, adenomyosis, apathy, auras, chills, cold, constipation, cramps, dehydration, diarrhea, distress, dizziness, drowsiness, endometriosis, fatigue, headache, heartburn, hiccups, hunger, ileus, labyrinthitis, lethargy, lightheadedness, migraines, nausea, overdosed, phobias, puberty, retching, tachycardia, thirst, vertigo, and vomiting. 
Cluster 2 (red) contains anger, anorexia, anxiety, bulimia, cardiomegaly, choking, cough, delirium, depressed, hypomania, hypotension, hysteria, insomnia, irritable, mania, neuropathy, numb, onycholysis, palpitations, shakiness, tinnitus, trauma, and worry. 
Cluster 3 (green) contains arthritis, burns, dermatitis, fibromyalgia, gastritis, hoarseness, hypertension, hyperthyroidism, hypothyroidism, inflammation, jaundice, ketonemia, ketosis, osteoarthritis, paralysis, pneumonia, rash, swallowing, and thyroiditis. 
Cluster 4 (purple) contains encephalitis and schizophrenia.
Additionally, we revealed similar ASEs that affiliate with the same cluster, such as `headache' and `migraines' (Fig. \ref{fig:graph}).

%---------------------------------------------
\section{Discussion and Conclusions}
\label{sec:discussion}
%---------------------------------------------

Obesity is a major health concern globally, affecting over 650 million individuals around the world \cite{haththotuwa2020worldwide}. 
GLP-1 RA used to treat obesity and T2D have demonstrated efficacy in reducing excess weight, improving comorbidities, and lowering blood sugar levels \cite{trujillo2021glp}. 
Because some GLP-1 RA were approved quite recently, their complete ASE profiles might remain under-reported or be overlooked in conventional studies. 
Clinical trials involving a few thousand human subjects are known to have difficulties in detecting ASEs that are rare or have a significant latency of development. 
Moreover, GLP-1 RA were originally developed to treat T2D, as opposed to obesity. 
To address this knowledge gap about ASEs, which poses risks for patients and physicians, we develop and present in this study a knowledge discovery approach that incorporates social media data and pharmacovigilance information to identify novel ASEs of GLP-1 RA.

To identify ASEs in texts, we applied a Named Entity Recognition (NER) model to analyze posts from $\mathbb{X}$ and Reddit that discuss GLP-1 RA, as well as academic papers that study GLP-1 RA. We also gathered ASEs along with their frequency from ChatGPT and manufacturers’ reports.
Many ASEs are included in all of our datasets, yet 16 out of 134 (12\%) are reported exclusively by academic studies, 14 (10\%) solely by manufacturers, and 21 (15\%) only on social media.
Social media provides a vast and diverse source of information, as demonstrated here, capturing a broader spectrum of user perspectives than other data sources, including those that may not be apparent in clinical research. 
The ability of our method to identify potentially novel ASEs, validated by academic studies, thereby supporting hypothesis \textit{H1}, has been demonstrated by focusing on ASEs present in both social media and academic papers (Group 2 in Fig. \ref{fig:venn}) but not reported by manufacturers. 
Our modeling approach was able to identify 53\% of established ASEs reported by manufacturers and ChatGPT (Fig. \ref{fig:overlap}). 
The ability to identify established ASEs provides further confidence in our modeling approach for exploring novel ASEs and supporting their novelty. 
Notably, established ASEs declared by manufacturers but not identified on social media have Very Rare frequency, as observed in Groups 4 and 7 (Fig. \ref{fig:venn}), and therefore are less likely to be reported on social media.

Furthermore, it is important to note ASEs reported very frequently on social media but not reported by manufacturers (Groups 2 and 5 in Fig. \ref{fig:venn}).
This discrepancy may stem from individuals using GLP-1 RA for weight loss, diverging from the drug’s original indication of treating diabetes.
Academic studies and manufacturers reporting the frequency of ASEs often rely on clinical trials, while social media offers much more data. Therefore, comparing the frequency of ASE mentions on social media to those reported in clinical trials was of great interest in the present study. In Fig. \ref{fig:drg_ase_35}, it is evident that the drug semaglutide (Ozempic) has the highest number of reported ASEs, followed by dulaglutide and liraglutide, respectively. However, as Ozempic is also the most-prescribed drug among those analyzed in this study, it is expected to have a higher frequency of reported ASEs. 
Our drug-ASE matrix analysis (Fig. \ref{fig:drg_ase_35}) highlights numerous prevalent ASEs based exclusively on social media reports. 
Additionally, comparing mentions of ASEs on social media in intervals of 14 days (Fig. \ref{fig:ASE_over_time}) to data from Google Trends (Fig. \ref{fig:Gtrends}), we found that the ASEs mentioned most frequently over time are nausea, pain, vomiting, and constipation.
To address RQ1, we identified ASEs solely based on raw social media discussions and constructed an ASE-ASE network (Fig. \ref{fig:graph}) $G$. 
In $G$, nodes represent ASEs, edges signify ASE co-occurrences in the same post, and the weight of an ASE-ASE edge represents the number of times this ASE co-occurrence pair was co-mentioned in the same post. 

Node clustering analysis of the ASE-ASE network revealed four clusters of ASEs, allowing a better understanding of their relationships:
Cluster 1 -- gastrointestinal distress; 
Cluster 2 -- emotional and mental strain syndromes; 
Cluster 3 -- somatic discomfort; and 
Cluster 4 -- neurological disorders. 
The presence of co-occurring ASEs is apparent (Fig \ref{fig:graph}), particularly with respect to Cluster 1 (light blue nodes), with strong and specific associations (thick network links/edges) among the common ASEs vomiting, nausea, migraines, fatigue, constipation, headache, and heartburn. 
This network analysis suggests that pairs of ASEs in Cluster 1 are frequently correlated with each other, and future work should examine in more detail the co-occurrence of these ASEs and compare these associations as reported by social media posts against manufacturers' reports and published literature.
Revealing groups of ASEs that commonly co-occur can highlight potential individual differences among GLP-1 RA users in terms of adverse effects experienced and provide insights into the interrelationships among different ASEs, potential mechanisms underlying adverse effects, and opportunities for personalized healthcare approaches that could utilize the complex, and sometimes rare, ASE-related relationships revealed by our analytical strategy.
 
Our analytical approach can potentially aid in the timely implementation of regulatory interventions, adjustments in treatment guidelines, and enhancement of patient safety measures. 
Moreover, monitoring fluctuations in the frequency and patterns of reported ASEs on social media need not be limited to GLP-1 RA.
The methods we use could enable detection of emerging issues or safety concerns related to any drug. 
Our comprehensive data analytics approach applied to large-scale social media data has substantial implications for drug safety assessment and for public health.
The integration of diverse data sources allows for more holistic discovery of adverse effects than standard pharmacovigilance strategies provide.
Here, the identified clusters contribute to a nuanced understanding of the therapeutic landscape of GLP-1 RA. 
Specifically, two important contributions of our study are that only 53\% of ASEs reported by drug manufacturers have been mentioned on social media, and moreover, 21 ASEs found on social media have not been reported by manufacturers. 
These findings call for additional studies.
Future work could further examine the association of drug popularity and awareness in popular culture with the sentiments expressed about the drug on social media. 
Additional research could use data from private discussions on social media to compare the sentiments expressed there with those voiced in public fora.
Our data-analytic knowledge discovery approach can be used for any drug to bring to light currently unidentified or under-reported side effects. 
Improving our understanding of the adverse effects of drugs in this way can serve to refine prescription recommendations and accordingly reduce  morbidity, thereby increasing positive outcomes for manufacturer, physician and patient alike.
Our approach paves the way for a new era of real-time pharmacovigilance, leveraging social media data to improve drug safety and public health outcomes.

%%===========================================================================================%%
%Bibliography
\bibliographystyle{unsrt}  
\bibliography{references}  
%%===========================================================================================%%

\newpage

%%===========================================================================================%%
\begin{appendices}
%%===========================================================================================%%

\section{
A list of GLP-1 receptor agonist adverse side effects (ASEs) identified in the datasets collected for this study.}
\label{app1:All_ASEs}
\begin{longtable}{llll}
\caption{A list of GLP-1 receptor agonist adverse side effects (ASEs) identified in the five datasets collected for this study.}
\\
\hline
ASE                                      & Frequency & Code & Legend                  \\ \hline
\endhead
\hline
\endfoot
\endlastfoot
Inflammation                             & 1376      & 1    & All                     \\
Hypoglycemia                             & 1006      & 1    & All                     \\
Depression                               & 981       & 1    & All                     \\
Diarrhea                                 & 951       & 1    & All                     \\
Nausea                                   & 820       & 1    & All                     \\
Increased Heart Rate                     & 736       & 1    & All                     \\
Microalbuminuria                         & 678       & 1    & All                     \\
Chills                                   & 446       & 1    & All                     \\
Vomiting                                 & 404       & 1    & All                     \\
Fatigue                                  & 390       & 1    & All                     \\
Constipation                             & 333       & 1    & All                     \\
Anxiety                                  & 320       & 1    & All                     \\
Headaches                                & 245       & 1    & All                     \\
Cramps                                   & 231       & 1    & All                     \\
Thyroid Issues                           & 185       & 1    & All                     \\
Gastritis                                & 168       & 1    & All                     \\
Rash                                     & 118       & 1    & All                     \\
Dizziness                                & 82        & 1    & All                     \\
Hypertrophy                              & 82        & 1    & All                     \\
Hypersensitivity                         & 80        & 1    & All                     \\
Dyspepsia                                & 79        & 1    & All                     \\
Swallowing Difficulty                    & 74        & 1    & All                     \\
Thirst                                   & 68        & 1    & All                     \\
Insomnia                                 & 65        & 1    & All                     \\
Stress                                   & 61        & 1    & All                     \\
Thrombosis                               & 44        & 1    & All                     \\
Otitis                                   & 35        & 1    & All                     \\
Jaundice                                 & 19        & 1    & All                     \\
Cholestasis                              & 10        & 1    & All                     \\
Cholelithiasis                           & 8         & 1    & All                     \\
Hidradenitis                             & 6         & 1    & All                     \\
Thromboembolism                          & 6         & 1    & All                     \\
Chest Tightness                          & 5         & 1    & All                     \\
Hyperglycemia                            & 5         & 1    & All                     \\
Polyuria                                 & 5         & 1    & All                     \\
Infarction                               & 729       & 2    & PubMed \& Social        \\
Ulcerative Wounds                        & 681       & 2    & PubMed \& Social        \\
Atherosclerosis                          & 567       & 2    & PubMed \& Social        \\
Neuropathy                               & 334       & 2    & PubMed \& Social        \\
Dehydration                              & 303       & 2    & PubMed \& Social        \\
Hypothyroidism                           & 122       & 2    & PubMed \& Social        \\
Hepatic Issues                           & 97        & 2    & PubMed \& Social        \\
Trauma                                   & 90        & 2    & PubMed \& Social        \\
Arthritis                                & 81        & 2    & PubMed \& Social        \\
Arrhythmias                              & 77        & 2    & PubMed \& Social        \\
Anorexia                                 & 69        & 2    & PubMed \& Social        \\
Dementia                                 & 68        & 2    & PubMed \& Social        \\
Ketosis                                  & 66        & 2    & PubMed \& Social        \\
Allodynia                                & 56        & 2    & PubMed \& Social        \\
Ketoacidosis                             & 54        & 2    & PubMed \& Social        \\
Osteoarthritis                           & 54        & 2    & PubMed \& Social        \\
Encephalopathy                           & 46        & 2    & PubMed \& Social        \\
Mania                                    & 42        & 2    & PubMed \& Social        \\
Hypotonia                                & 41        & 2    & PubMed \& Social        \\
Palpitations                             & 41        & 2    & PubMed \& Social        \\
Cholangitis                              & 40        & 2    & PubMed \& Social        \\
Urticaria                                & 40        & 2    & PubMed \& Social        \\
Cardiomyopathy                           & 38        & 2    & PubMed \& Social        \\
Acidosis                                 & 36        & 2    & PubMed \& Social        \\
Delusion                                 & 31        & 2    & PubMed \& Social        \\
Erythema                                 & 30        & 2    & PubMed \& Social        \\
Tachycardia                              & 30        & 2    & PubMed \& Social        \\
Compulsive Behavior                      & 25        & 2    & PubMed \& Social        \\
Gangrene                                 & 23        & 2    & PubMed \& Social        \\
Pneumonia                                & 21        & 2    & PubMed \& Social        \\
Osteomyelitis                            & 19        & 2    & PubMed \& Social        \\
Amnesia                                  & 16        & 2    & PubMed \& Social        \\
Lethargy                                 & 13        & 2    & PubMed \& Social        \\
Malnutrition                             & 11        & 2    & PubMed \& Social        \\
Fractures                                & 8         & 2    & PubMed \& Social        \\
Retching                                 & 8         & 2    & PubMed \& Social        \\
Bradycardia                              & 7         & 2    & PubMed \& Social        \\
Fasciitis                                & 6         & 2    & PubMed \& Social        \\
Glycosuria                               & 6         & 2    & PubMed \& Social        \\
Hypercholesterolemia                     & 6         & 2    & PubMed \& Social        \\
Neurotoxicity                            & 6         & 2    & PubMed \& Social        \\
Bezoar                                   & 230       & 3    & PubMed                  \\
Atheroma                                 & 91        & 3    & PubMed                  \\
Hypoxia                                  & 68        & 3    & PubMed                  \\
Nephrotoxicity                           & 67        & 3    & PubMed                  \\
Pruritus (Itching)                       & 67        & 3    & PubMed                  \\
Stiffness                                & 42        & 3    & PubMed                  \\
Hyperphagia                              & 32        & 3    & PubMed                  \\
Cardiotoxicity                           & 26        & 3    & PubMed                  \\
Ischemia                                 & 24        & 3    & PubMed                  \\
Polyneuropathies                         & 23        & 3    & PubMed                  \\
Paresthesia                              & 20        & 3    & PubMed                  \\
Hyperplasia                              & 18        & 3    & PubMed                  \\
Arteriosclerosis                         & 8         & 3    & PubMed                  \\
Demyelination                            & 8         & 3    & PubMed                  \\
Dyskinesia                               & 8         & 3    & PubMed                  \\
Hyperemia                                & 8         & 3    & PubMed                  \\
Increase in Amylase                      & NA        & 4    & Manufacturers \& GPT    \\
Serious Allergic Reaction                & NA        & 4    & Manufacturers \& GPT    \\
Thyroid C-Cell Tumors                    & NA        & 4    & Manufacturers \& GPT    \\
Injection Site Nodule                    & 14.5      & 4    & Manufacturers \& GPT    \\
Sinus Tachycardia                        & 11.5      & 4    & Manufacturers \& GPT    \\
Hair Loss                                & 6         & 4    & Manufacturers \& GPT    \\
First-Degree Atrioventricular (AV) Block & 4.6       & 4    & Manufacturers \& GPT    \\
Elevated Serum Lipase                    & 2.1       & 4    & Manufacturers \& GPT    \\
Dysgeusia                                & 1.7       & 4    & Manufacturers \& GPT    \\
High Calcitonin                          & 1.2       & 4    & Manufacturers \& GPT    \\
Breast Cancer                            & 0.7       & 4    & Manufacturers \& GPT    \\
Colorectal Neoplasms                     & 0.6       & 4    & Manufacturers \& GPT    \\
Papillary Thyroid Cancer                 & 0.2       & 4    & Manufacturers \& GPT    \\
Elevated Serum Amylase                   & 0.1       & 4    & Manufacturers \& GPT    \\
Irritability                             & 58        & 5    & Social only             \\
Burns                                    & 46        & 5    & Social only             \\
Numbness                                 & 45        & 5    & Social only             \\
Hypogonadism                             & 31        & 5    & Social only             \\
Cough                                    & 26        & 5    & Social only             \\
Paralysis                                & 24        & 5    & Social only             \\
Anger                                    & 22        & 5    & Social only             \\
Hoarseness                               & 22        & 5    & Social only             \\
Bulimia                                  & 19        & 5    & Social only             \\
Suicidal                                 & 18        & 5    & Social only             \\
Anhedonia                                & 12        & 5    & Social only             \\
Frustration                              & 12        & 5    & Social only             \\
Onycholysis                              & 12        & 5    & Social only             \\
Endometriosis                            & 11        & 5    & Social only             \\
Ketonemia                                & 6         & 5    & Social only             \\
Apathy                                   & 4         & 5    & Social only             \\
Aura                                     & 4         & 5    & Social only             \\
Hirsutism                                & 4         & 5    & Social only             \\
Infertility                              & 4         & 5    & Social only             \\
Narcolepsy                               & 4         & 5    & Social only             \\
Snoring                                  & 4         & 5    & Social only             \\
Eructation                               & 35        & 6    & Manufacturers \& Social \\
Bloating                                 & 30        & 6    & Manufacturers \& Social \\
Back Pain                                & 12        & 6    & Manufacturers \& Social \\
Hepatotoxicity                           & 6         & 6    & Manufacturers \& Social \\
Syncope                                  & 4.6       & 6    & Manufacturers \& Social \\
Appendicitis                             & 3         & 6    & Manufacturers \& Social \\
Blurred Vision                           & 15        & 7    & Pubmed \& manufacturers \\ \hline
\label{tbl:ases}
\end{longtable}

\end{appendices}

\end{document}